\documentclass[a4paper,11pt]{article}
\usepackage{pos}
\usepackage[utf8]{inputenc}
\usepackage{textgreek}
\usepackage{color}
\usepackage{bm}

\title{Physics-informed neural networks for angular-momentum conservation in computational relativistic spin hydrodynamics}

\author*[a]{Hidefumi Matsuda}
\author[a,b]{Koichi Hattori}
\author[b]{Koichi Murase}

\affiliation[a]{Zhejiang Institute of Modern Physics, Department of Physics, Zhejiang University,\\
Hangzhou, 310027, China}

\affiliation[b]{Research Center for Nuclear Physics (RCNP), Osaka University,\\
Osaka 567-0047, Japan}

\emailAdd{da.matsu.00.bbb.kobe@gmail.com}
\emailAdd{koichi.hattori@zju.edu.cn}
\emailAdd{phys.murase@gmail.com}

\abstract{
Theoretical developments in relativistic spin hydrodynamics, which describes the macroscopic transport of spin angular momentum alongside other fundamental conserved quantities, have progressed rapidly since the experimental observation of the global spin polarization of $\Lambda$ hyperons in relativistic heavy-ion collision experiments. However, numerical simulations of relativistic spin hydrodynamics remain largely unaddressed due to computational challenges, particularly the accurate numerical conservation of total angular momentum.
In this work, we propose the use of physics-informed neural networks (PINNs) for computational relativistic spin hydrodynamics. As a concrete application, we consider a rotating fluid confined within a cylindrical container.
We show that angular-momentum conservation can be accurately achieved in the PINNs-based numerical framework. Furthermore, we investigate the spin-orbit conversion induced by the rotational viscous effect, which is the intrinsic dissipative process of relativistic spin hydrodynamics. 
Our analysis numerically identifies the mismatch between the transverse thermal vorticity and the spin potential as the driving mechanism of the spin-orbit conversion.
}

\FullConference{
26th International Spin Physics Symposium (SPIN 2025)\\
21-26 September 2025\\
Qingdao, Shandong, China
}

\begin{document}
\maketitle

\section{Introduction}
The experimental discovery of spin polarization and alignment~\cite{STAR:2017ckg,ALICE:2019aid} in relativistic heavy-ion collisions has motivated the development of relativistic spin hydrodynamics~\cite{Montenegro:2017rbu,Florkowski:2017ruc,Hattori:2019lfp}, which describes the macroscopic transport of the spin degrees of freedom carried by quarks and gluons. Despite considerable theoretical progress, numerical simulations of relativistic spin hydrodynamics remain extremely scarce~\cite{Singh:2024cub,Sapna:2025yss}. Specifically, no existing simulations incorporate the mutual conversion between spin and orbital angular momentum, which is expected to play a role in interpreting spin-related observables in heavy-ion collisions.
Conventional finite-volume methods (FVMs) for hydrodynamics solve continuity equations while ensuring that the conservation laws are implemented free from discretization errors. This guarantee, however, applies only to the conserved quantities associated with the equations explicitly solved.
As a result, the implementation of angular momentum conservation in FVMs poses a challenge; since it is often not independent of energy and momentum conservation laws, the corresponding equations are typically not solved entirely explicitly. Therefore, it is worthwhile to explore alternative numerical methods that can ensure the conservation of angular momentum alongside other conserved quantities in a consistent manner.

In this work, we propose the use of Physics-Informed Neural Networks (PINNs)~\cite{raissi2019physics} as a novel numerical approach to computational relativistic spin hydrodynamics. We apply this approach to a rotating fluid confined within a cylindrical container. We first verify that the PINNs-based approach accurately preserves total angular momentum. Following this verification, we numerically study how the rotational viscous effect leads to the spin-orbit conversion. We use the mostly positive metric $\eta_{\mu\nu}=\mathop{\mathrm{diag}}(-1,1,1,1)$.  Further details on the theoretical
formulation and numerical implementation can be found in our full paper~\cite{Matsuda:2025whj}.

\section{Second-order Relativistic Spin Hydrodynamics}\label{Sec2}
We begin by outlining the formulation of a second-order viscous relativistic spin hydrodynamics in $(3+1)$ dimensions~\cite{Matsuda:2025whj}.
We treat the spin density $S^{\mu\nu}$ as an additional thermodynamic variable that relaxes slowly compared to other non-hydrodynamic modes.
The fundamental thermodynamic relation for the internal energy density $e$ is expressed as $de = Tds +
\omega_{\mu\nu} dS^{\mu\nu}$, where $T$ and $s$ are the temperature and the entropy density, respectively, and the spin potential $\omega_{\mu\nu}$ is the thermodynamic conjugate of $S^{\mu\nu}$.
Both $S^{\mu\nu}$ and $\omega_{\mu\nu}$ are antisymmetric and spatial, $u_\mu S^{\mu\nu}=u_\mu\omega^{\mu\nu}=0$, with $u^\mu$ being the flow velocity.  The energy--momentum (EM) tensor $\Theta^{\mu\nu}$ and the spin current tensor $\Sigma^{\mu\nu\rho}$ satisfy the conservation laws of
energy--momentum and total angular momentum: $\nabla_\mu \Theta^{\mu\nu}=0$ and $\nabla_\mu \Sigma^{\mu\nu\rho}=-2\Theta^{[\nu\rho]}$.
We adopt the Landau frame, $\Theta^{(\mu\nu)}u_\nu=-eu^\mu$, and the pseudo-gauge in which the spin tensor $\Sigma^{\mu\nu\rho}$ 
is totally antisymmetric with respect to all its indices.
The tensor decompositions of the EM and spin tensors are given by
\begin{align}
\Theta^{\mu\nu} = e u^\mu u^\nu + P \Delta^{\mu\nu} + \phi^{\mu\nu}\ , \qquad
\Sigma^{\mu\alpha\beta} = u^\mu S^{\alpha\beta} - u^\alpha S^{\mu\beta} + u^\beta S^{\mu\alpha}\ ,
\end{align}
with the pressure $P$ and the spatial projector $\Delta^{\mu\nu} = \eta^{\mu\nu} + u^\mu u^\nu$.
Here we neglect bulk and shear viscosities.
The antisymmetric part of the EM tensor, the couple-stress tensor $\phi^{\mu\nu}$, is a dissipative correction that obeys the relaxation equation derived based on the M\"uller--Israel--Stewart approach~\cite{Romatschke:2017ejr},
\begin{align}
\tau_\phi
\Delta^\mu_{\ \alpha}
\Delta^\nu_{\ \beta}
D\phi^{\alpha\beta}
= - 2\gamma \rho^{\mu\nu}
  - \phi^{\mu\nu}
  - \frac{\tau_\phi}{2}(\theta + \beta^{-1}D\beta) \phi^{\mu\nu}\ ,\label{Eq:IS}
\end{align}
where $D = u_\mu \nabla^\mu$ is the convective derivative, and $\gamma$ and $\tau_\phi$ are the rotational viscosity and its relaxation time, respectively. The rotation-rate mismatch $\rho^{\mu\nu} = \beta^{-1} \varpi^{\mu \nu}_\perp - 2\omega^{\mu\nu}$ is the difference between the transversely projected thermal vorticity,
$\varpi^{\mu \nu}_\perp \equiv \Delta^{\mu \alpha} \Delta^{\nu \beta} \partial_{[\alpha} \beta_{\beta]}$, and the spin potential.
In the small $\tau_\phi$ limit with fixed $\gamma$,
$\phi^{\mu\nu}$ reduces to the first-order correction $-2\gamma\rho^{\mu\nu}$.
In Sec.~\ref{Sec3}, these equations are applied to the system under consideration and then solved numerically.

\section{PINNs-based computational framework}
We solve the hydrodynamic equations using the framework of Physics-Informed Neural Networks (PINNs)~\cite{raissi2019physics}. 
In this approach, a neural network is trained to approximate the solution of partial differential equations by minimizing a loss function 
that encodes the governing equations and boundary conditions. 
A distinctive feature of PINNs is the flexibility to incorporate various physical constraints into the training process. 
With use of this advantage, we define the total loss function to include the conservation of the total angular momentum, defined as
\begin{align}
\mathcal{L}_{\text{total}} = \mathcal{L}_{\text{hydro}} + \mathcal{L}_{\text{BC}} + \mathcal{L}_{\text{local}}^{(\text{AM})} + \mathcal{L}_{\text{global}}^{(\text{AM})}\ ,
\end{align}
where each term penalizes the violation of a specific condition: $\mathcal{L}_{\text{hydro}}$ for the hydrodynamic equations, 
$\mathcal{L}_{\text{BC}}$ for the boundary conditions, and $\mathcal{L}_{\text{local}}^{(\text{AM})}$ and $\mathcal{L}_{\text{global}}^{(\text{AM})}$ 
for the local and global conservation of angular momentum, defined as
\begin{align}
\nabla_\mu J^{\mu xy}(x)=0\ ,\qquad
\int d^3x\, J^{txy}(t,\bm{x})=\text{const.}\ ,
\end{align}
respectively.
Here, we assume vanishing angular-momentum flux through the
boundary of the space. We only consider the total angular momentum in the $z$ direction, $J^{txy} = x\Theta^{ty} - y\Theta^{tx} + \Sigma^{txy}$, because we restrict the fluid dynamics to a two dimensional disk in the next section.

\section{Numerical Demonstration of Mutual Spin--Orbit Conversion}\label{Sec3}
We investigate the dynamical conversion between orbital and spin angular momentum during fluid evolution according to relativistic spin hydrodynamics. We consider a rotating fluid confined
within a cylindrical geometry of radius $R$ within the cylindrical coordinates $(r,\theta,z)$.  We assume
translational invariance along the $z$ direction and neglect parity-odd tensor components with respect to $z$, 
effectively reducing the dynamics to a two-dimensional disk in the transverse plane. 
Furthermore, rotational symmetry is assumed, simplifying the problem to $(1+1)$-dimensional evolution in time $t$ and radial coordinate $r$. 
In the following, all physical quantities are normalized by the cylinder radius $R$.

Under the aforementioned conditions, we solve the hydrodynamic equations derived in Sec.~\ref{Sec2} for the independent variables $\{e, u^r, u^\theta, S^z, \phi^{r\theta}\}$:
\begin{align}
De &= - \frac{4}{3} e \theta + u_\nu \partial_\mu \phi^{\mu\nu}\ ,\label{Eq:hydro1}\\
\frac{4}{3}e D u^{ r    } &= - \frac{1}{3}u^{ r    } D e - \frac{1}{3}\nabla^{ r    } e - \Delta^r{}_\nu \nabla_\mu \phi^{\mu \nu}\ ,\label{Eq:hydro2}\\
\frac{4}{3}e D u^{\theta} &= - \frac{1}{3}u^{\theta} D e - \Delta^{\theta}{}_\nu \nabla_\mu \phi^{\mu \nu}\ ,\label{Eq:hydro3}\\
\partial_t S^z &= -2 r \phi^{r \theta}\ ,\label{Eq:hydro4}\\
\tau_\phi
\Delta^r{}_{\alpha}
\Delta^\theta{}_{\beta}
D\phi^{\alpha\beta}
  &=- 2\gamma \rho^{r \theta}
  - \phi^{r \theta}
 - \frac{2\tau_\phi}{3} (\nabla_\alpha u^\alpha) \phi^{r \theta}\ .\label{Eq:hydro5}
 \end{align}
Regarding the boundary conditions, we impose vanishing conditions on all fields at the center ($r=0$).
At the outer boundary ($r=R=1$), we set $u^r =
\phi^{r\theta} = S^z = 0$, and additionally impose Neumann-type conditions
on $u^r$ and $\phi^{r\theta}$.  
To study the mutual conversion process between orbital and spin angular momentum, 
we consider two distinct non-equilibrium configurations as initial conditions, by introducing
detivation to the global equilibrium state with $e=1$ and all other fields vanishing:
\begin{itemize}
\item \textbf{Initial Condition~F:} We introduce a non-vanishinig azimuthal
  velocity profile, $u^\theta = 0.2 \sin^4(\pi r)$, on top of
  the global equilibrium state. This explores how a
  rotating fluid generates spin polarization through the Barnett effect.
\item \textbf{Initial Condition~S:} Conversely, we introduce a non-vanishing
  spin polarization profile, $S^z = 0.2 \sin^4(\pi r)$, on top of
  the global equilibrium state. This explores
  how initial spin polarization induces macroscopic fluid rotation through the
  Einstein--de Haas effect. 
\end{itemize}
Note that, to ensure the system of hydrodynamic equations is closed and 
solvable, we supplement them with thermodynamic relations (i.e., the equation
of state, heat capacity, and spin susceptibility) for a conformal, massless,
two-flavor quark--gluon gas.

We use a Multi-Layer Perceptron (MLP) as the base architecture for the PINNs. The network consists of three hidden layers with 250 units each. We use the hyperbolic tangent ($\tanh$) as the activation function for all layers and initialize the weights using the Xavier method. To minimize the loss function, we use the Adam optimizer with an initial learning rate between $10^{-3}$ and $10^{-5}$. Automatic differentiation is handled by the PyTorch autograd engine, 
and the batch size is set to 150,000.
As training proceeds, we observe a rapid decrease of the loss function in
the early stages, followed by a slower reduction until the loss function eventually reaches a plateau. This plateau indicates that the training has sufficiently converged. All results presented below are obtained from this converged model.

\begin{figure*}[tp]
\begin{minipage}{0.5\textwidth}
    \centering
    \includegraphics[width=\textwidth]{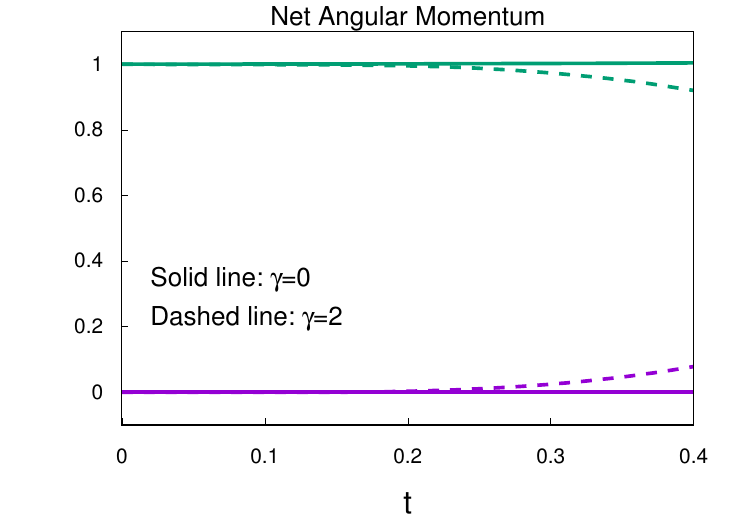}
\end{minipage}
\begin{minipage}{0.5\textwidth}
    \centering
    \includegraphics[width=\textwidth]{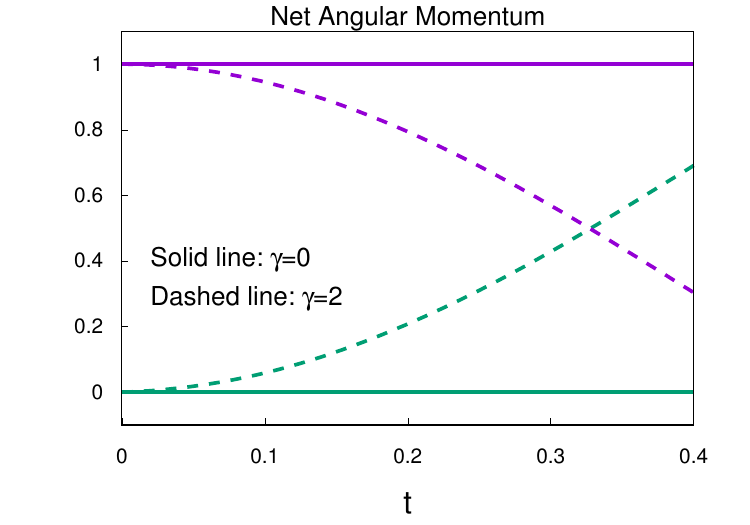}
\end{minipage}
\caption{
Time evolution of the net orbital and spin angular momentum for $\gamma=0$ and $\gamma=2$. The left and right panels correspond to Initial Conditions~F and~S, respectively. Each quantity is normalized by the total angular momentum (the sum of net orbital and spin components) at the initial time $t=0$. In both panels, the green and purple curves represent the net orbital and net spin angular momentum, respectively.}
\label{Fig:ang_HIC}
\end{figure*}

\begin{figure*}[tp]
\begin{minipage}{0.33\textwidth}
    \centering
    \includegraphics[width=\textwidth]{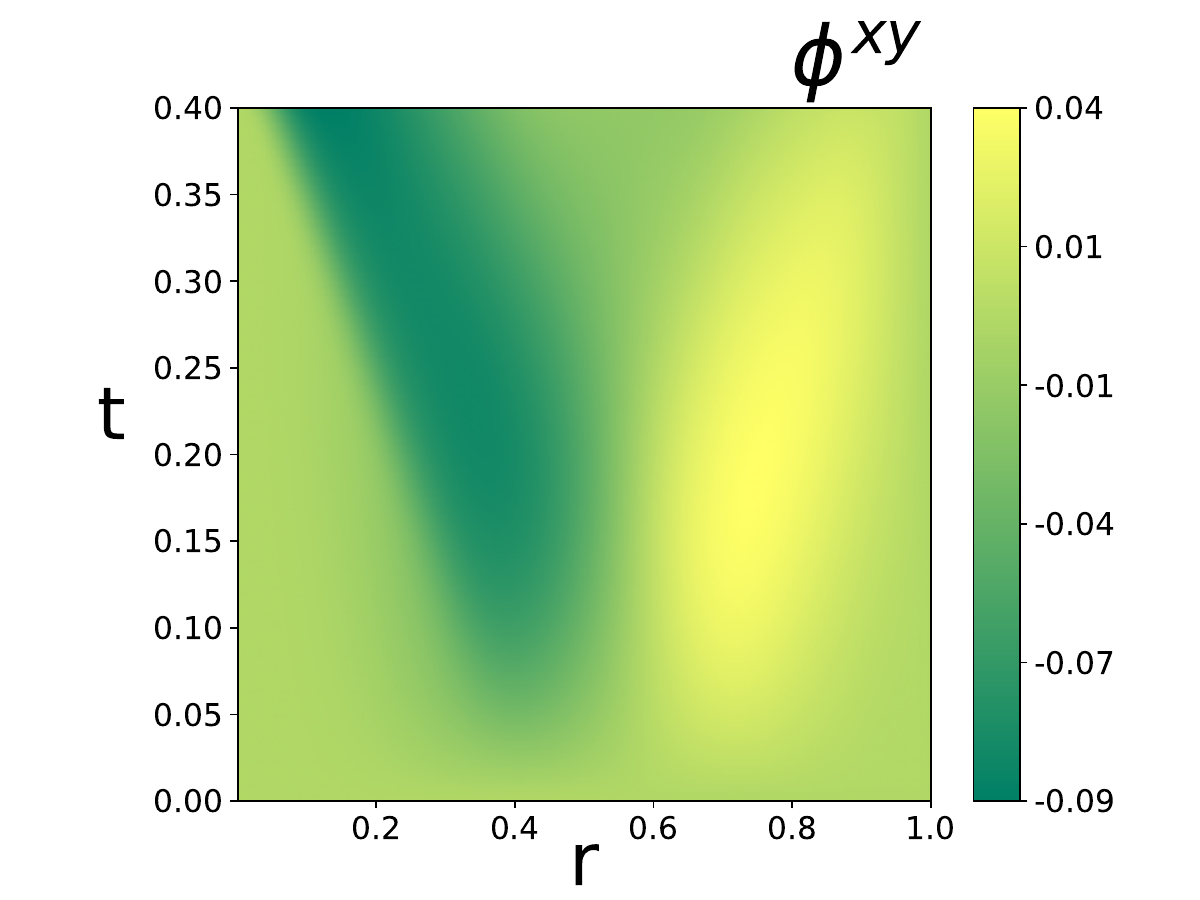}
\end{minipage}
\begin{minipage}{0.33\textwidth}
    \centering
    \includegraphics[width=\textwidth]{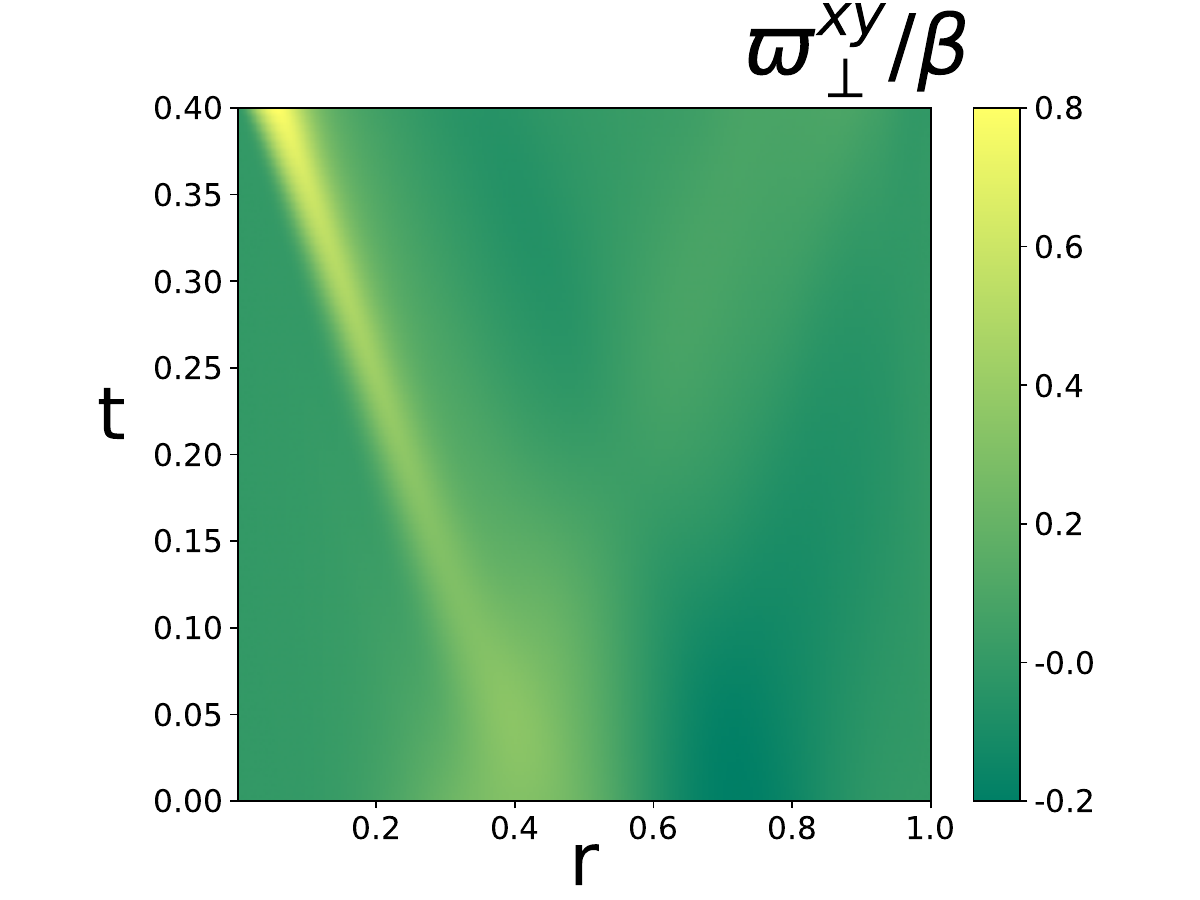}
\end{minipage}
\begin{minipage}{0.33\textwidth}
    \centering
    \includegraphics[width=\textwidth]{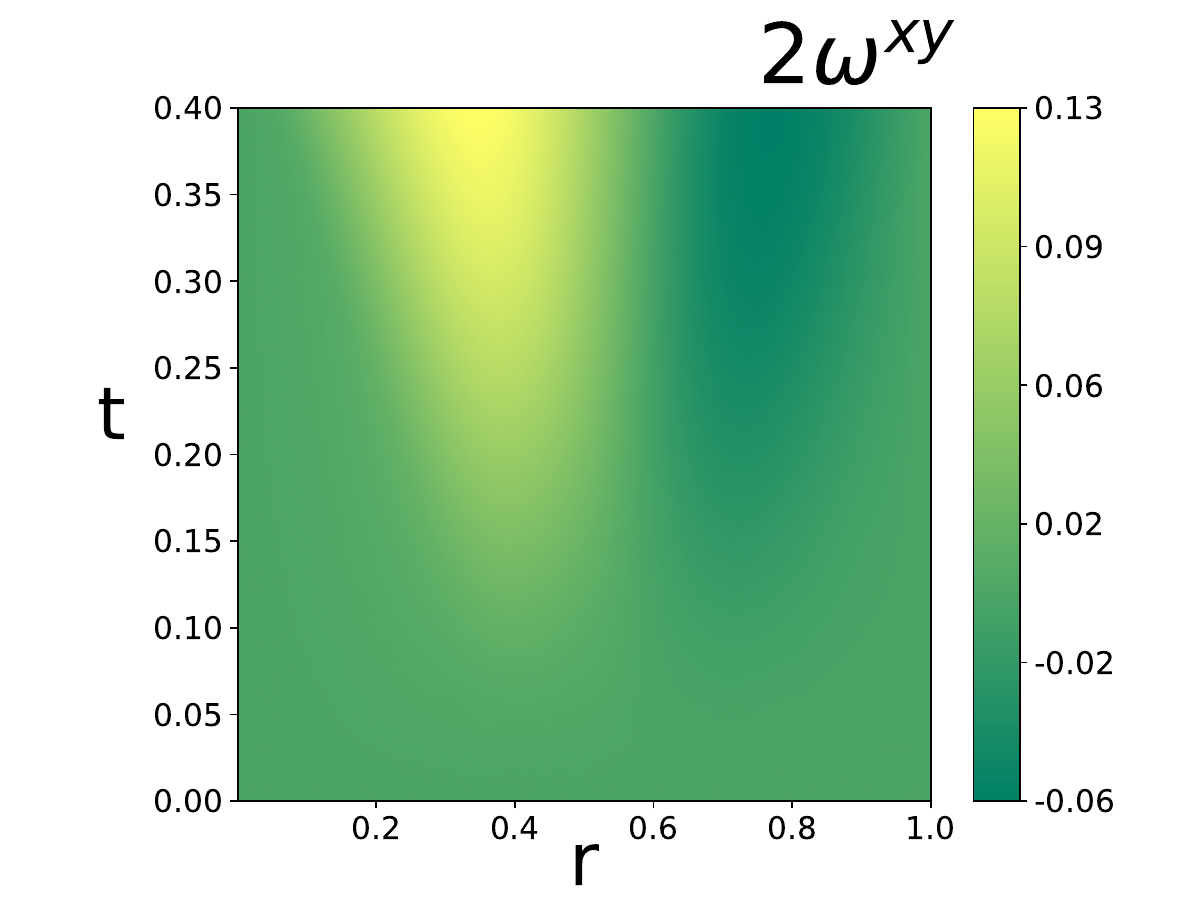}
\end{minipage}
\begin{minipage}{0.33\textwidth}
    \centering
    \includegraphics[width=\textwidth]{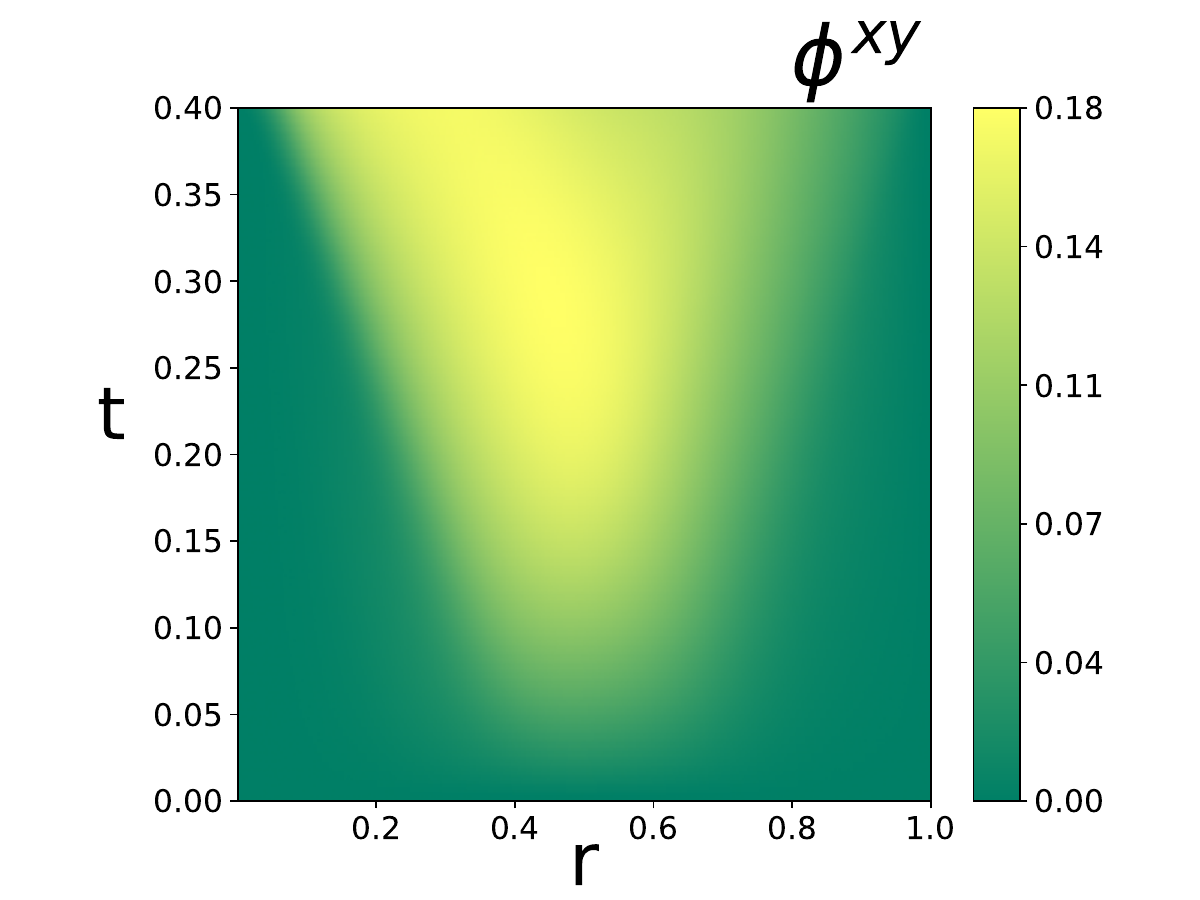}
\end{minipage}
\begin{minipage}{0.33\textwidth}
    \centering
    \includegraphics[width=\textwidth]{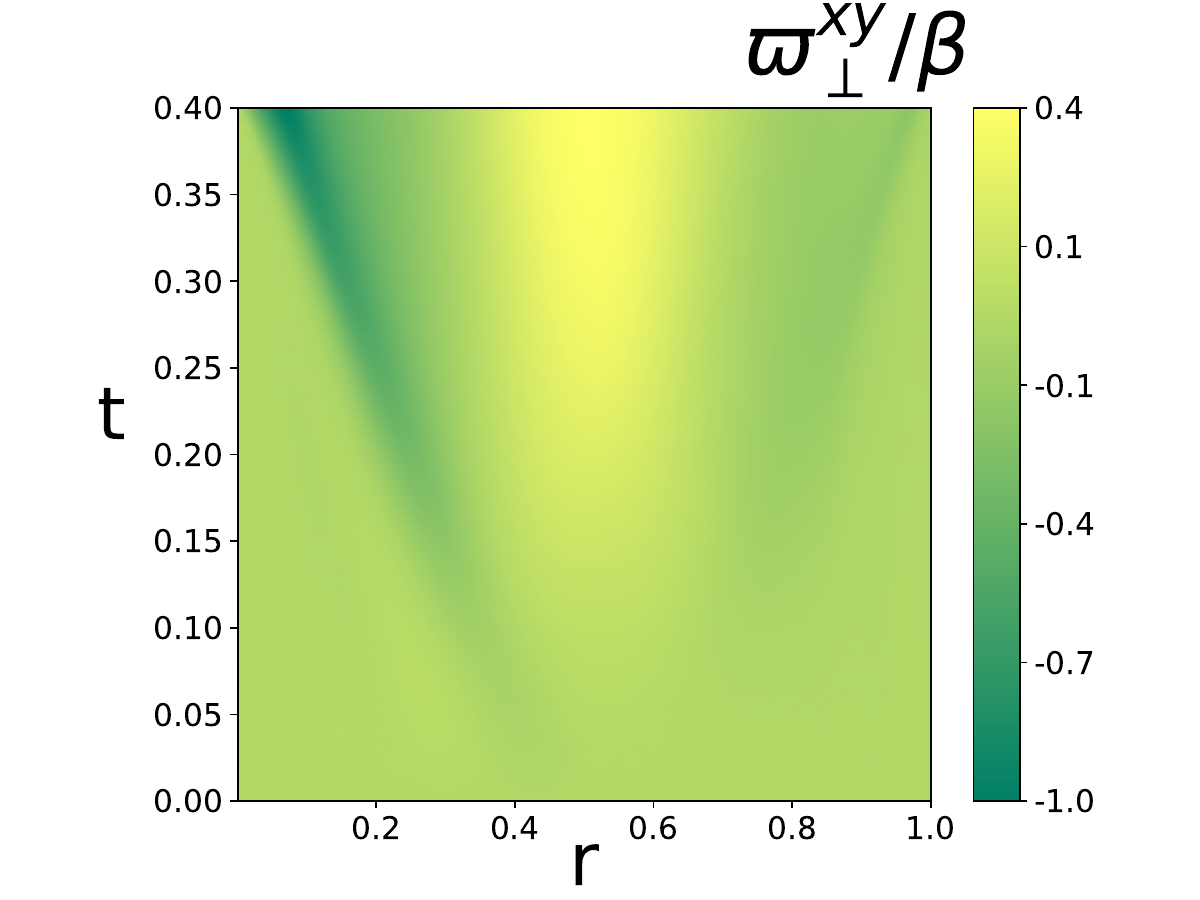}
\end{minipage}
\begin{minipage}{0.33\textwidth}
    \centering
    \includegraphics[width=\textwidth]{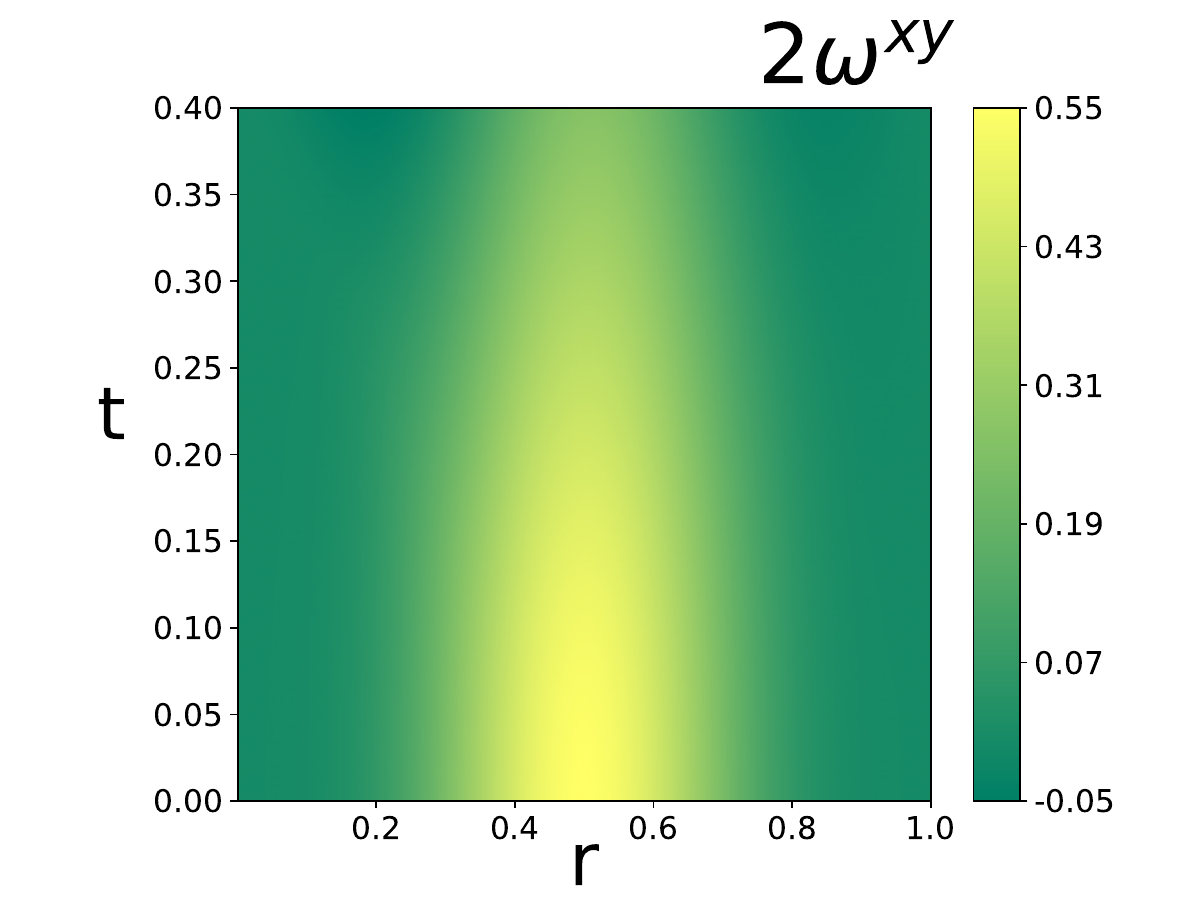}
\end{minipage}
\caption{
Spacetime evolution of the $xy$ components for the couple-stress tensor, transverse thermal vorticity, and spin potential at $\gamma=2$. 
The three columns, from left to right, represent these respective quantities. Initial Condition~F is displayed in the top row, while the results for Initial Condition~S are shown in the bottom row.
}
\label{Fig:phi_HIC}
\end{figure*}

Figure~\ref{Fig:phi_HIC} shows the $xy$ components of 
the couple-stress tensor $\phi^{xy}$, the transverse thermal vorticity $\varpi^{xy}_\perp$, and the spin potential $\omega^{xy}$, where the difference between the latter two is the rotational-rate mismatch $\rho^{xy}$. For Initial Condition~F (top row of Fig.~\ref{Fig:phi_HIC}), $\varpi^{xy}_\perp$ is positive inside the cylinder and negative outside at $t=0$, while $\omega^{xy}$ is negligible. Consequently, the early-time evolution of $\phi^{xy}$ closely follows the spatial structure of $\varpi^{xy}_\perp$, as seen in the two left panels. The later-time distribution of $\omega^{xy}$ is explained by the fact that $\phi^{xy}$ acts as the sole source term for the spin density in Eq.~\eqref{Eq:hydro4}, with the spin density being proportional to the spin potential in our current setting. 

The bottom row of Fig.~\ref{Fig:phi_HIC} shows the results for Initial Condition~S. Although $\phi^{xy}$ vanishes initially, a pronounced peak appears around $r=0.5$ at $t \approx 0.02$. This peak correlates well with the spatial distribution of the spin potential $\omega^{xy}$, which is also localized near $r=0.5$. This is because, at early times, the initial growth of $\phi^{xy}$ is driven by the large mismatch between the finite $\omega^{xy}$ and vanishing $\varpi^{xy}_\perp$. Subsequently, $\varpi^{xy}_\perp$ evolves so as to reduce the mismatch with the spin potential.

\section{Summary}
In this work, we have developed a novel computational framework based on Physics-Informed Neural Networks (PINNs) for relativistic spin hydrodynamics. Our approach addresses the challenge of ensuring total angular momentum conservation.
By investigating two complementary physical scenarios representing the Barnett and Einstein--de Haas effects, we have numerically demonstrated the mutual conversion between orbital and spin angular momentum. Our analysis reveals that the spin-orbit conversion is driven by the rotation-rate mismatch, which is the difference between the transverse thermal vorticity and the spin potential. This mismatch enhances the couple-stress tensor, which acts as the mediator for the angular momentum exchange. These findings not only validate the PINNs-based methodology but also provide insights into the dissipative mechanisms in relativistic spin hydrodynamics.

\section*{Acknowledgments}
We acknowledge that numerical calculations were performed on Supercomputer
Yukawa-21 at Yukawa Institute for Theoretical Physics (YITP) at Kyoto
University.  This work is supported by the JSPS KAKENHI under Grant
Nos.~JP22H01216, JP23H05439, and~JP23K13102.

\bibliographystyle{unsrt}
\bibliography{main}       

\end{document}